\documentclass[twocolumn,showpacs,preprintnumbers,amssymb,epsfig,prl,aps]{revtex4}

\usepackage{graphicx}
\usepackage{dcolumn}
\usepackage{bm}

\begin{document}

\preprint{}

\title{Electric field effect modulation of transition temperature, mobile
carrier density and in-plane penetration depth in NdBa$_{2}$Cu$_{3}$O$%
_{7-\delta }$ thin films}
\author{D. Matthey}
\altaffiliation[Now at ]{Department of Physics and Astronomy, University of Aarhus, Ny Munkegade,
Building 1520, DK-8000 Aarhus C}
\email{daniel@inano.dk}
\author{N. Reyren}
\author{J.-M. Triscone}
\affiliation{DPMC, University of Geneva, 24 quai Ernest-Ansermet, 1211 Geneva 4,
Switzerland. }
\author{T. Schneider}
\affiliation{Physikinstitut, University of Zurich, Winterthurerstrasse 190, 8057 Zurich,
Switzerland. }
\date{\today }

\begin{abstract}
We explore the relationship between the critical temperature,
$T_{c}$, the mobile areal carrier density, $n_{2\text{D}}$, and the
zero temperature magnetic in-plane penetration depth, $\lambda
_{ab}(0)$, in very thin underdoped NdBa$_{2}$Cu$_{3}$O$_{7-\delta }$
films near the superconductor to insulator transition using the
electric field effect technique. Having established consistency with
a Kosterlitz-Thouless transition we observe that $T_{KT}$ depends
linearly on $n_{2\text{D}}$, the signature of a quantum
superconductor to insulator transition in two dimensions with
$z\overline{\nu }=1$, where $z$ is the dynamic and $\overline{\nu }$
the critical exponent of the in-plane correlation length.
\end{abstract}

 \maketitle

The electronic properties of high $T_{c}$ superconductors are
critically determined by the density of mobile holes as illustrated
by their generic phase diagram in the temperature dopant
concentration plane. Understanding the physics of this phase
diagram, in particular at the  two $T=0$ edges of the
superconducting dome, has emerged as one of the critical questions
in the field of cuprate superconductors.  In the underdoped regime,
where a superconductor to insulator transition occurs, there is an
empirical linear relation (the so-called Uemura relation) between
$T_{c}$ and $\lambda _{ab}^{-2}(0)$ \cite{uemuraprl89}, where
$\lambda _{ab}$ is the in-plane London penetration depth. If, in the
underdoped limit, a two dimensional (2D) quantum superconductor to
insulator (QSI) transition occurs, such a relation between $T_c$ and
$1/\lambda^{2}(0)$ is expected \cite{schneiderbook}. This however
necessitates the occurrence of a 3D to 2D crossover as the
underdoped limit is approached with a diverging anisotropy $ \gamma
=\lambda _{c}/\lambda _{ab}$. In the case of a 3D QSI-transition,
one would expect $T_{c}\propto \lambda _{a,b,c}\left( 0\right)
^{-2z/\left( z+1\right) }$, where $z$ is the dynamic critical
exponent of the transition \cite{schneiderbook}. Recent penetration
depth
measurements on YBa$_{2}$Cu$_{3}$O$_{6+x}$ single crystals \cite%
{hosseini04,broun05} (where $\gamma $ apparently saturates at low
doping levels) and films \cite{zuevprl05} suggest a 3D-transition
with $2z/\left( z+1\right) \simeq 1$ \cite{ts05comment}, a result
different from the Uemura relation. Furthermore, an empirical
relation involving the normal state conductivity and extending up to
optimum doping was recently proposed by Homes {\it et al.}
\cite{homesnat04}. However, all these studies on the relationship
between $T_{c}$ and $\lambda _{a,b,c}\left( 0\right) $ stem from
samples where the doping level was modified by chemical substitution
\cite{uemuraprl89,schneiderbook,hosseini04,broun05,zuevprl05,ts05comment,tallonprb03,homesnat04}.

In this letter we use the electrostatic field effect to tune the
mobile areal carrier density, $n_{2\text{D}}$, in ultrathin,
strongly underdoped NdBa$_{2}$Cu$_{3}$O$_{7-\delta }$ (NBCO) films
and explore the intrinsic relation between $T_{c}$, $n_{2\text{D}}$
and $\lambda _{ab}^{-2}(0)$ close to the superconductor to insulator
transition. The electrostatic field effect technique, which allows
the carrier density in a given sample to be changed without
affecting the chemical composition, thus avoiding sample to sample
variations and substitution induced disorder,  appears to be an
ideal method to elucidate the intrinsic relationships between the
key physical parameters \cite{ahnnat03}.

In a classical oxide field effect configuration, an electric field
is applied across a gate dielectric in a heterostructure made of
an oxide channel and a dielectric layer. The density of carriers
is modified at the interface between the oxide channel and the
dielectric, changing the electronic properties of the oxide
channel. This technique allows us to induce in thin oxide
superconducting films rather large $T_{c}$ modulations
\cite{mattheyapl03}. In the 3-4 unit cell thick films used in
this study, $T_{c}$ modulations of more than $10$ K have been
achieved, corresponding to induced charge densities of the order
of $0.7\cdot 10^{14}$ charges/cm$^{2}$.

\begin{figure}[tbp]
\centering
\includegraphics[angle=0,width=8.6cm]{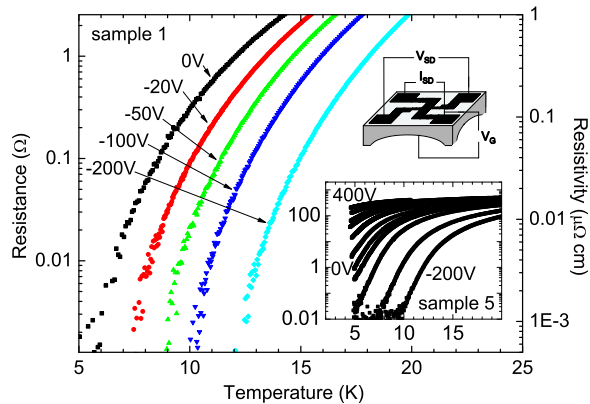}
\caption{Resistance versus temperature for a 3-4
unit cell thick NBCO film (sample 1) for different applied gate
voltages; the right scale is the corresponding resistivity. The lower inset shows the resistance as a function of
temperature for sample 5, which is also 3-4 unit cell thick, for
different applied gate voltages (see text for details). The upper
inset shows a sketch of the field effect device, based on a thinned
STO single crystal gate insulator.} \label{fig1}
\end{figure}

The field effect device used in this
study, described in Ref. \onlinecite{mattheyapl03}, is based on a SrTiO$_3$ (STO) single
crystal gate dielectric, the substrate itself. Due to its large
low temperature dielectric constant, $\varepsilon$, and large achievable
 polarizations, STO is a particularly
 interesting gate insulator. Field effect devices using a STO thin film gate
 insulator have thus been studied
extensively \cite{mannhartsst96}. Here, thin $100$ $\mu $m thick, or
thinned $500$ $\mu $m commercial substrates have been used. A sketch
of the thinned device is shown in the upper inset of Fig.\ref{fig1}.
The superconducting NBCO thin films are first grown by off-axis
magnetron sputtering onto $(001)$ (bare or etched) STO substrates
heated to about $730^{\circ }$C. During cooling, a typical O$_{2}$
pressure of $670$ Torr is used to obtain optimally doped films. To
control the initial doping level of the films, the oxygen cooling
pressure is modified and lowered down to $5$ mTorr for the most
underdoped films. Gold is  sputtered in-situ at room temperature to
improve the contact resistances and the whole structure is protected
by an amorphous NBCO layer also deposited in-situ. X-ray diffraction
allowed us to determine precisely the film thickness down to three
unit cell thick films. After deposition, the samples are
photolithographically patterned using ion milling and a gold
electrode is deposited on the backside of the samples, facing the
central part of the superconducting path. The measured path is $600$
(length)$\times500$ (width) $\mu $m$^2$. The patterning process does
not affect $T_c$. All the thin films presented here are 3-4 unit
cell thick.

Fig.\ref{fig1} shows resistance versus temperature for sample 1 in
the tail of the transitions, for different voltages applied across
the gate dielectric: $0$, $-20$, $-50$, $-100$ and $-200$V. Negative
voltages correspond to a negative potential applied to the gate and
a positive one to the superconducting path, resulting in hole doping
of the oxide channel and, as expected, in a ``shift'' to higher
temperatures of the resistive transition. Resistance measurements
are performed using a standard four point technique while a voltage,
$V_{g}$, is applied to the gate. During measurements, gate leakage
currents were kept below a few nA, while the current used for
resistance measurements was typically between $1$ and $10\mu $A. The
lower inset of Fig.\ref{fig1} shows resistance versus temperature
for sample $5$ and for gate voltages of $400$, $200$, $100$, $50$,
$20$, $10$, $0$, $-5$, $-20$ , $-50$, $-100$, and $-200$ V. At
$4.2$K, a large increase in resistance is observed while applying
positive gate voltages, effectively reducing the hole concentration.

\begin{figure}[tbp]
\centering
\includegraphics[angle=0,width=8.6cm]{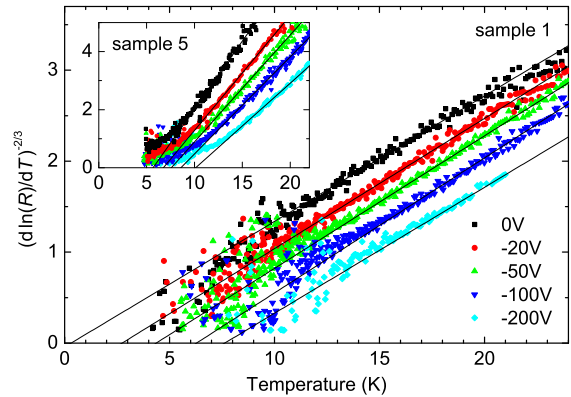}
\caption{$\left( \text{d}\ln R/\text{d}T\right) ^{-2/3}$ {\em vs.}
$T$ for samples 1 and 5 at various gate voltages. The solid lines
indicate the consistency with the linear KT relationship.
$T_{c}=T_{KT}$ is determined by the condition $\left( \text{d}\ln
R/\text{d}T\right) ^{-2/3}=0$ at $T_{KT}$.} \label{fig2}
\end{figure}

To estimate $T_{c}$, we explore the consistency of our resistivity
data with the expected behavior  for a Kosterlitz-Thouless (KT)
transition, namely  $\rho =\rho _{0}\exp (-bt^{-1/2})$, where
$t=T/T_{c}-1$ and $\rho _{0}$ and $b$ are material dependent but
temperature independent parameters \cite{schneiderbook}.
Accordingly, consistency with the KT-scenario is established when
the plot $\left( \text{d}\ln R/\text{d}T\right) ^{-2/3}$ {\em vs.}
$t$ exhibits, near $t=0$, linear behavior and $T_{c}=T_{KT}$ is
determined by the condition $\left( \text{d}\ln R/\text{d}T\right)
^{-2/3}=0$ at $T_{KT}$.
A glance at Fig.\ref{fig2} reveals that the characteristic
KT-behavior is well confirmed in an intermediate regime. In sample
1, it is the noise level which limits the accessible regime, while
in sample $5$ a rounded transition appears to occur, a signature
that the correlation length divergence is limited by the size of the
homogenous domains. Indeed, above $T_{KT}$ the correlation length
increases exponentially as $\xi =\xi _{0}\exp (bt^{-1/2})$. An
estimate of the size of the homogeneous regions for sample 5 for
$T\simeq 7.5$K, $T_{KT}=5.67$K and $b=2.63$ (values appropriate for
sample 5 at $V=0$), leads to $\xi \left( 7.5\text{K}\right) /\xi
_{0}\simeq 100$. Remarkably enough, with $\xi _{0}$ ranging from
$10$ to $100$\AA, the size of the homogeneous domains exceeds
$1000$\AA .

\begin{figure}[tbp]
\centering
\includegraphics[angle=0,width=8.6cm]{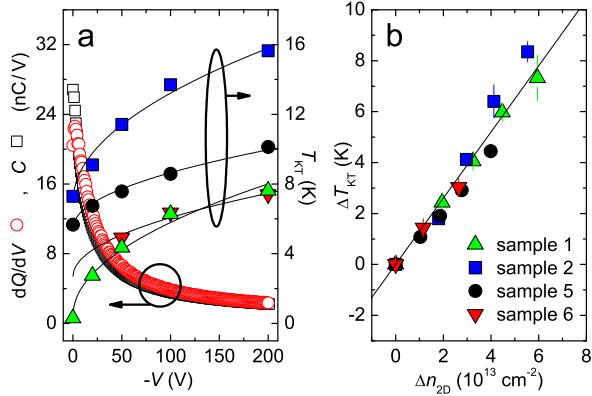}
\caption{ (a) Right scale, $T_{KT}$ versus gate voltage $V$. The
solid lines are $T_{KT}\left( V\right) \simeq T_{KT}\left( 0\right)
+a\left\vert V\right\vert^{1/2} $ with $a=0.54$, 0.62, 0.32,  0.34
for samples 1, 2, 5
and 6.
Left scale, $\text{d}Q/\text{d}V$ versus $V$
for a 100$\mu$m thick STO measured at 4.2 K using an electrometer (open circles), $C(V)$ measured using an LCR-meter (open squares). (b) $\Delta T_{KT}$ \textit{vs.} $\Delta n_{2\text{D}}$
for all the samples. The solid line is Eq.(\ref{eq1}).}
\label{fig3}
\end{figure}

The consistency with a KT-transition discussed above enables us to
estimate $T_{c}=T_{KT}$ and determine its gate-voltage dependence.
The resulting $T_{KT}$ estimates for various samples versus  gate
voltage $V$ are plotted in Fig.\ref{fig3}a. As can be seen,
$T_{KT}\left( V\right) $ is nonlinear and, as indicated by the
respective solid curves, $T_{KT}\left( V\right) \simeq T_{KT}\left(
0\right) +a\left\vert V\right\vert ^{1/2}$.

To correlate the field-induced $T_{KT}$ modulations to the
field-induced areal carrier density, $n_{2\text{D}}$, we measured
the accumulated charge $Q $. Since the dielectric constant of STO
depends on temperature and applied electric field, we measured the
field and temperature dependence of the capacitance
\cite{christenprb94}. The induced areal charge density, $\sigma
=e\Delta n_{2\text{D}}$, at a given temperature was measured in
terms of the capacitance of the device as a function of the applied
gate voltage, using an LCR meter (Agilent 4284A) with an
auto-balancing bridge method, and/or by measuring the charge flow
during loading using an electrometer (Keithley 6514). By ramping the
voltage across the dielectric, the LCR-meter measures the
\textquotedblleft local\textquotedblright\ capacitance $C(V)$ at a
given voltage while the electrometer measures $\text{d}Q$ for a
given change in voltage \footnote{$\text{d}Q$ is obtained for a
$\text{d}V$ of 1V. Also $C(V)$ was measured with voltage modulation
amplitudes between 0.1 and 1V. }. The two measurements agree
quantitatively as can be seen on Fig.\ref{fig3}a where
$\text{d}Q/\text{d}V$ and $C(V)$ ($\text{d}Q=C(V)\text{d}V$),
measured at $4.2$K, are plotted versus $V$ for a $100\mu $m thick
STO substrate with two parallel $20$ mm$^{2}$ square gold
electrodes. The corresponding field induced charge density $\sigma
(V)$ at a given voltage and temperature is then obtained from
$\sigma (V)=1/S\int_{0}^{V}C(V)\text{d}V$, where $S$ is the surface
of the gate contact. This quantitative estimate of $\sigma $ allows
a change in $T_{KT}$ to be related to a change in the areal carrier
density $n_{2\text{D}}=\sigma /e$. The result shown in
Fig.\ref{fig3}b reveals the intrinsic linear relationship
\begin{equation}
\centering\Delta T_{KT}=1.3\ 10^{-13}\Delta n_{2\text{D}},
\label{eq1}
\end{equation}
where $\Delta T_{KT}=T_{KT}(V) -T_{KT}(0)$ \footnote{Except for sample 6 where $\Delta T_{KT}=T_{KT}(V) -T_{KT}(-50\text{V})$.}
and $\Delta n_{2\text{D}}$ is expressed in $\text{cm}^{-2}$. This is
a novel and central result of our paper.

Together with the quantum
counterpart of the Nelson-Kosterlitz relation \cite{nelson},
$T_{KT}\propto $ $\lambda _{ab}^{-2}(T_{KT})$, it implies that
$T_{KT}$ , $\lambda _{ab}(0)$ and $n_{2\text{D}}$ are universally related by
\begin{equation}
\centering T_{KT}\propto n_{2\text{D}}\propto \frac{1}{\lambda
_{ab}^{2}(0)}, \label{eq2}
\end{equation}
with non-universal factors of proportionality. This relationship
also confirms the theoretical predictions for a 2D-QSI transition.
Indeed, the scaling theory of quantum critical phenomena
\cite{schneiderbook,kim91} predicts that close to a QSI transition
$T_{c}$ scales as $n_{2\text{D}}^{z\overline{\nu }}$, where $z$ is
the dynamic and $\overline{\nu }$ the critical exponent of the zero
temperature in-plane correlation length, $\xi _{ab}\propto
n_{2\text{D}}^{-\overline{\nu }}$. Thus, Eq.(\ref{eq2}) reveals
that $z\overline{\nu }\simeq 1$, the signature of a QSI transition
in 2D \cite{herbut0,herbut01}. Since close to a QSI transition
$\lambda _{ab}(0)$ scales as $1/\lambda _{ab}^{2}(0)\propto
n_{2\text{D}}^{\overline{\nu }\left( D+z-2)\right) }$ ($D$ is the
system dimensionality) \cite{schneiderbook,kim91}, it follows that
$\Delta T_{KT}\propto \Delta n_{2\text{D}}$ does not only uncover a
2D-QSI transition with $z\overline{\nu }\simeq 1$, but also implies
that $T_{KT}$ , $\lambda _{ab}(0)$ and $n_{2\text{D}}$ are
universally related by Eq.(\ref{eq2}). However, this relationship
differs drastically from $T_{c}\propto 1/\lambda _{a,b,c}(0)$,
derived from penetration depth measurements on YBa$_{2}$Cu$_{3}
$O$_{6+x}$ single crystals \cite{hosseini04,broun05} and films
\cite{zuevprl05}, where $T_{c}$ was reduced by chemical
substitution.

\begin{figure}[tbp]
\centering
\includegraphics[angle=0,width=8.6cm]{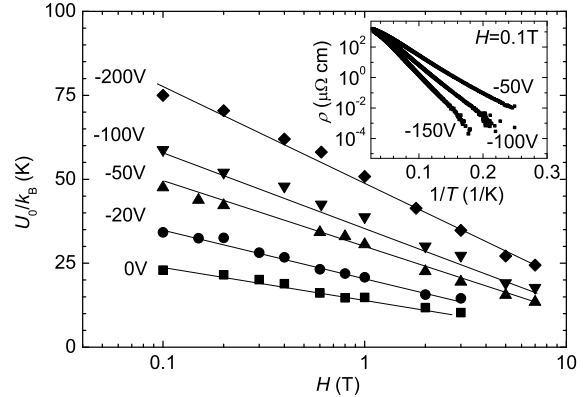}
\caption{$U_0/k_B$ \textit{vs.} $H$ for different
applied voltages for sample 1. A linear relation between $U_0$ and
$\log(H)$ is observed with a slope which depends on the applied
electric field. Inset: $\log (\rho)$ as a function of $1/T$ for
sample 2 and for three different electrical fields at $0.1$T. The
zero temperature activation energy $U_0$ at a given applied
electrical field is obtained from the slope of the Arrhenius plot.}
\label{figure4}
\end{figure}

To substantiate our main result further, we explore the temperature,
electric and magnetic field dependence of the resistive transition,
extracting the activation energy, $U$, for vortex motion in the
liquid vortex phase. In the vortex 2D-limit the resistance in a
field is activated with activation energy proportional to
$-$ln$\left( H\right) $. The measurements have been performed with
the magnetic field applied perpendicular to the $ab$-plane, ramping
the temperature slowly and measuring the sample resistance at a
given magnetic and electric field. Inset of Fig.\ref{figure4} shows
Arrhenius plots of the resistivity, $\log (\rho )$ as a function of
$1/T$, for sample 2 and for three different electrical fields at
$H=0.1$T. The observed linear relation between $\log (\rho )$ and
$1/T$ singles out thermally activated flux flow, where $\rho
(H,T)=\rho _{n}\exp (-U(H,T)/k_{B}T)$, with $U(H,T)=2U_{0}(H)\left(
1-T/T_{c}\right) $ for $T\sim T_{c}$ \cite{palstraprb90}. As can be
seen, $U$ becomes larger for electrical fields raising the number of
holes and $T_{c}$. We note that an electrostatic modulation of the
activation energies was also observed in reference
\onlinecite{walkenhorst}. From such Arrhenius plots we estimated
$U_{0}(H)/k_{B}$ for sample 1 at different applied voltages and
different magnetic fields. As can be seen in Fig.\ref{figure4}, at
a given voltage, we observe between $0.1$ and $7$ T the
characteristic 2D logarithmic field dependence $U_{0}(H)=-\alpha \ln
(H)+\beta $, in agreement with previous measurements on thin films,
superlattices, and the highly anisotropic
kappa-(BEDT-TTF)$_{2}$Cu(NCS)$_{2}$ \cite{brunnerprl91,fischer92,
white93, suzuki94, friemel96, zhang00}. Note that in 3D $U\propto
H^{-1/2}$ is expected \cite{blatter94}. Furthermore we find that
$T_{c}=T_{KT}$ is proportional to $U_{0}(H)/k_{B}$ for every value
of the magnetic field. Since $U$ is proportional to $1/\lambda
_{ab}^{2}$ in $D=2$ (and $D=3$) \cite{feigelmanphc90,blatter94},
this proportionality is also consistent with our main result
$T_{c}\propto n_{2D}\propto 1/\lambda _{ab}^{2}(0)$.

In summary, we explored the relationships between $T_{c}$, the
mobile areal carrier density, $n_{2\text{D}}$, and the zero
temperature in-plane penetration depth, $\lambda _{ab}(0)$, for very
thin underdoped NBCO films near the superconductor to insulator
transition by means of the electric field effect technique. Together
with the observed behavior of the resistive transition we
established remarkable consistency with 2D critical behavior and a
quantum superconductor to insulator transition, characterized by a
linear relationship between $T_{c}$, $n_{2\text{D}}$ and $1/\lambda
_{ab}^{2}(0)$. This result also implies that isotope or pressure
effects on $T_{c} $, $n_{2\text{D}}$ and $\lambda _{ab}(0)$ are
related accordingly.

We would like to thank P. Martinoli and S. Gariglio for useful
discussions and M. Dawber for a careful reading of the manuscript.
This work was supported by the Swiss National Science Foundation
through the National Center of Competence in Research, ``Materials
with Novel Electronic Properties, MaNEP'' and division II, New
Energy and Industrial Technology Development Organization (NEDO)
of Japan, and ESF (Thiox).

\end{document}